\documentclass[%
 reprint,
 amsmath,amssymb,
 aps,
]{revtex4-2}

\usepackage{graphicx}
\usepackage{dcolumn}
\usepackage{bm}
\usepackage{hyperref}
\usepackage{comment}

\usepackage{subfigure}
\usepackage{color}

\begin{document}

\preprint{APS/123-QED}

\title{Exploring asymmetric multifractal cross-correlations of price-volatility and asymmetric volatility dynamics in cryptocurrency markets}

\author{Shinji Kakinaka}
 \altaffiliation[Corresponding author, ]{Department of Applied Mathematics and Physics, Graduate School of Informatics, Kyoto University, Japan.}
\author{Ken Umeno}%

\date{\today}

\begin{abstract}
Asymmetric relationship between price and volatility is a prominent feature of the financial market time series. This paper explores the price-volatility nexus in cryptocurrency markets and investigates the presence of asymmetric volatility effect between uptrend (bull) and downtrend (bear) regimes. The conventional GARCH-class models have shown that in cryptocurrency markets, asymmetric reactions of volatility to returns differ from those of other traditional financial assets. We address this issue from a viewpoint of fractal analysis, which can cover the nonlinear interactions and the self-similarity properties widely acknowledged in the field of econophysics. The asymmetric cross-correlations between price and volatility for Bitcoin (BTC), Ethereum (ETH), Ripple (XRP), and Litecoin (LTC) during the period from June 1, 2016 to December 28, 2020 are investigated using the MF-ADCCA method and quantified via the asymmetric DCCA coefficient. The approaches take into account the nonlinearity and asymmetric multifractal scaling properties, providing new insights in investigating the relationships in a dynamical way. We find that cross-correlations are stronger in downtrend markets than in uptrend markets for maturing BTC and ETH. In contrast, for XRP and LTC, inverted reactions are present where cross-correlations are stronger in uptrend markets.
\begin{description}
\item[Keywords]
price-volatility cross-correlations; asymmetric volatility; cryptocurrency markets;\\
MF-ADCCA; asymmetric DCCA coefficients.
\end{description}
\end{abstract}

\maketitle

\section{Introduction}
Bitcoin (BTC) was first released in 2009 by a pseudonymous publisher Satoshi Nakamoto~\cite{Nakamoto2019}, who introduced a decentralized and pseudo-anonymous system supported by the block-chain technology. Differently from other financial assets, the unique system of BTC allows anonymous electric payment to put into practical use while avoiding duplicated transactions. It was not until 2013, when the BTC market underwent two price bubbles within the same year, that BTC began to be recognized by specialists and experts, increasing its awareness to a wider area. The year 2017 was a decisive year for BTC where the skyrocketing increase of market prices took notice and attracted the attention of a wide community including active online traders and academic researchers. During the same period, competing cryptocurrencies such as Ethereum (ETH), Ripple (XRP), Litecoin (LTC) have emerged and grown rapidly dominating more than half of the cryptocurrency market capitalization in recent years, indicating that analyzing minor coins has also become important. Researchers have discussed whether cryptocurrencies should be classified as currencies, financial assets, an expedient medium of exchange, or a technological-based product~\cite{Lo2014, Blau2017, Yermack2015, Polasik2015, White2020}, but they have not come to a complete conclusion. Moreover, BTC is less correlated with conventional assets, commodities, and the U.S. dollar, making it useful as a diversified investment for hedging purposes~\cite{Bouri2017hedge}. The market capitalization is on its rising trend and has surpassed a staggering 58 billion dollars at the end of 2020. Given these idiosyncratic characteristics, modeling the fundamental features of BTC, along with other cryptocurrencies, plays a crucial role in various types of financial analyses.

Numerous studies have shown that financial assets are incredibly complex and have nonlinear dynamic systems with scaling properties, power-law distributions with fat-tails, long-range dependencies, and volatility clustering. Such stylized facts are seen in cryptocurrency markets as well~\cite{Bariviera2017, Drozdz2018, alvarez2018long, kakinaka2020characterizing}. Begu\v{s}i\'{c} et al.~(2018)~\cite{Begui2018} study the distribution of BTC returns and find empirical evidence of slowly decaying tails of power-law behavior with $2 < \alpha < 2.5$. This suggests that BTC returns, in addition to being more volatile, exhibit heavier tails than stocks, where the exponent of stock returns is known to be around 3. Kakinaka and Umeno (2020)~\cite{kakinaka2020characterizing} find that the bulk part of the distribution is well described by the stable distribution.
The BTC returns do not follow a random walk behavior and the market is significantly inefficient, but the efficiency becomes higher in recent periods~\cite{Urquhart2016} and the market heads to maturity~\cite{Drozdz2018}. Although some studies suggest that market efficiency holds for certain periods, the returns do not generally satisfy the efficient market hypothesis (EMH)~\cite{Bariviera2017ECL, Tiwari2018, zhang2018inefficiency}. Jiang et al.~(2018)~\cite{jiang2018time} investigate how the Hurst exponent varies through a rolling window approach and conclude that the BTC market has a high degree of inefficiency over time. Bariviera (2017)~\cite{Bariviera2017ECL} uses a dynamical approach of detrended fluctuation analysis (DFA) proposed by Peng et al.~(1994)~\cite{Peng1994mosaic}, which can be applied to non-stationary data and provides more reliable estimates of the Hurst exponent compared to the traditional rescaled range analysis~\cite{hurst1957suggested}. Bariviera finds long-memory and captures the underlying dynamic process generating the prices and volatility in the BTC market.

Multifractal detrended fluctuation analysis (MFDFA) is a development of the DFA algorithm, overcoming the limitations of DFA by quantitatively describing multifractal structures in terms of the generalized scaling exponents~\cite{kantelhardt2002multifractal}. The analysis of financial series using the MFDFA has lead to a breakthrough in the field of econophysics as an effective approach to detect inefficiency, multifractality, and long-memory in a nonlinear way. Many studies have applied the MFDFA and found evidence that cryptocurrency markets have strong multifractality originating from correlations and fat-tails~\cite{da2018multifractal, takaishi2018statistical, al2018efficiency, shrestha2019multifractal, stavroyiannis2019high}. The generalization of the DFA to cross-correlations between bivariate series is known as the detrended cross-correlation analysis (DCCA) introduced by Podobnik and Stanley~(2008)~\cite{podobnik2008detrended}. Its extension to multifractality, the multifractal detrended cross-correlation analysis (MFDCCA, MFDXA, MF-X-DFA)~\cite{zhou2008multifractal}, is developed and implemented for the empirical studies of cryptocurrencies, stock prices, and crude oil markets~\cite{zhang2018inefficiency, zhang2018multifractal, el2019bitcoin, ghazani2020multifractal}. This method is a combination of the MFDFA and the DCCA, thus describing multifractal characteristics of cross-correlated non-stationary series. A different generalization of the DFA, the asymmetric DFA (A-DFA), was proposed by Alvarez-Ramirez et al.~(2009)~\cite{alvarez2009dfa} since financial assets may show different behavior in reaction to trends. The multifractal version was later proposed, namely, the asymmetric MFDFA (A-MFDFA)~\cite{cao2013asymmetric, lee2017asymmetric}, and the further extension to cross-correlations is known as the multifractal asymmetric DCCA (MF-ADCCA)~\cite{cao2014detrended}. The method of MF-ADCCA is a versatile tool that takes into account both the asymmetric structure and the multifractal scaling properties between the two series. An empirical study by Garjardo et al.~(2018)~\cite{gajardo2018does} applies the MF-ADCCA to price behaviors of BTC and leading conventional currencies, suggesting the presence and asymmetry of cross-correlations between them. Using the same approach, Kristjanpoller and Bouri~(2019)~\cite{kristjanpoller2019asymmetric} find evidence of asymmetric multifractality between the main cryptocurrencies and the world currencies. These studies show that the MF-ADCCA approach is powerful for uncovering complex systems in cryptocurrency markets~\footnote{For more technical information and further discussion relevant to the detrending based multifractal methods, see~\cite{wkatorek2020multiscale}}.

One of the fundamental issues in financial markets is the behavior of the relationship between price and volatility. The cross-correlations between international stock returns of highly developed economies fluctuate strongly with time, and increase in periods of high market volatility~\cite{Solnik1996}. In addition, it is known that the conditional variance of equity returns are more affected by negative news compared to positive news~\cite{black1976studies}. In this sense, negative returns increase the volatility by more than positive returns due to the leverage effect, also known as the asymmetric volatility effect~\cite{baur2018asymmetric}. This is related to the trading of informed and uninformed investors posing different impacts on the return process. Specifically, uninformed traders lead to higher serial correlations in returns that make the volatility increase, whereas informed traders suggest no autocorrelation~\cite{avramov2006impact}.
Many studies have traditionally applied the asymmetric Generalized Autoregressive Conditional Heteroscedasticity (GARCH) models to analyze the asymmetric reactions of volatility to returns. Baur and Dimpfl~(2018)~\cite{baur2018asymmetric} use the TGRACH model and find an intriguing aspect of cryptocurrency price behavior differently from other traditional assets--- the presence of inverse-asymmetric volatility effect. In other words, positive shocks increase the volatility by more than negative shocks, where speculative investments made by uninformed noise traders are dominant after positive shocks. Cheikh et al.~(2020)~\cite{cheikh2020asymmetric} employ a flexible model of ST-GARCH and also report similar results of a positive return-volatility relationship.

Although the anomalies in volatility dynamics justify the application of GARCH class models~\cite{katsiampa2017volatility}, they focus more on the linear correlations of returns and volatility fluctuations without the scaling properties~\cite{ghazani2020multifractal}. Applying the DFA based analysis could be advantageous to accomplish a more essential understanding of the dynamics of price and volatility because it accounts for the widely acknowledged nonlinearity and scaling properties. In addition, the method does not require the specification of a statistical model, so it is easy to implement and can directly calculate the scaling exponents.

This paper employs the MF-ADCCA approach to investigate the asymmetric multifractal properties of cross-correlations between price fluctuations and realized volatility fluctuations in cryptocurrency markets, which is the first contribution to the literature. Our empirical findings reveal traces of asymmetric multifractality and its dependence on the directions of price movements. Next, we address the financial issue of asymmetric volatility effect, crucial for financial investors and regulators. We propose to use a framework of fractal analysis, which differs from the conventional GARCH-class models where how one-time price shock influences volatility is analyzed. This is the second contribution to the literature. We show that the asymmetric volatility effect can be examined by the assessment of asymmetric cross-correlations between bull and bear markets over various time scales from short to long time periods. The levels of cross-correlations are quantitatively investigated for each time scale relying on the asymmetric DCCA coefficient~\cite{cao2018multifractal} based on the idea of the DCCA coefficient~\cite{zebende2011dcca, podobnik2011statistical}. We find the presence of different reactions of volatility to price in major cryptocurrencies, where price-volatility is more strongly cross-correlated under negative market trends compared to positive market trends. However, for the relatively minor ones, cross-correlations are stronger under positive market trends, implying that the inverse-asymmetric volatility effect is present. Our findings not only provide new insights into the nature of the price-volatility dynamics, but also contribute to other relevant issues such as volatility spillovers, speculative trading, and the maturity of cryptocurrency markets.

The rest of this paper is organized as follows. Section 2 describes the data used in the analysis. Section 3 explains the two nonlinear dynamical methods used in the analysis, the MF-ADCCA approach and the asymmetric DCCA coefficient, to further investigate volatility dynamics. Section 4 present the results and discussions of the empirical analysis. Section 5 concludes.

\section{Data}
In this study, we use cryptocurrency price data traded on \url{https://poloniex.com/}~\footnote{We take advantage of Poloniex that the exchange provides high frequency data without missing data throughout the investigated period. Although Poloniex certainly may not be one of the most known exchanges, it provides transactions over 100 active cryptocurrencies. In the exchange, the markets such as BTC, ETH, XRP, and LTC have enough liquidity so that market analysis can be conducted without encountering zero values of intraday returns.} for four major coins of BTC, ETH, XRP, and LTC all against Tether (USDT), which is a cryptocurrency designed to maintain the same value as the US dollar. We analyze the period starting from June 1st, 2016, and ending on December 28th, 2020. This period includes the cryptocurrency boom at the end of 2018 and the crash at the beginning of 2019. Recently the prices of many cryptocurrencies are increasing, for instance, the Bitcoin price marked a record-breaker of over \$27,200 on December 28, 2020.

\begin{figure*}[ht]
    \begin{center}
        \subfigure[BTC]{\includegraphics[width=0.45\linewidth]{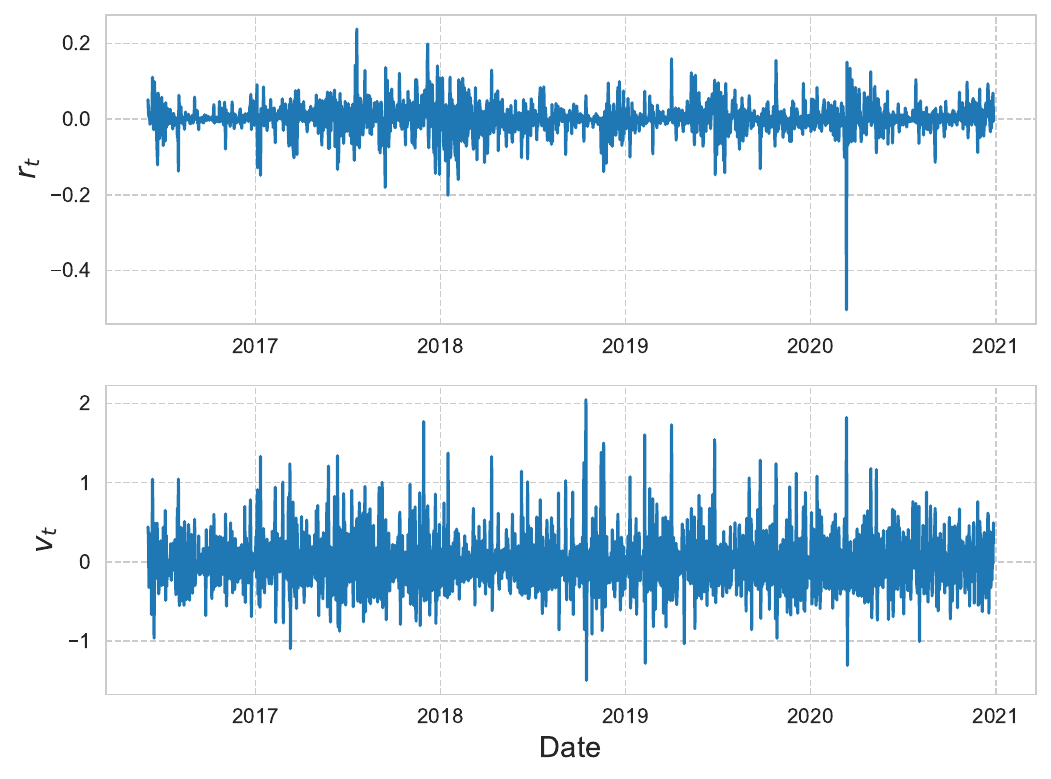}\label{fig:figBTC}}
        \subfigure[ETH]{\includegraphics[width=0.45\linewidth]{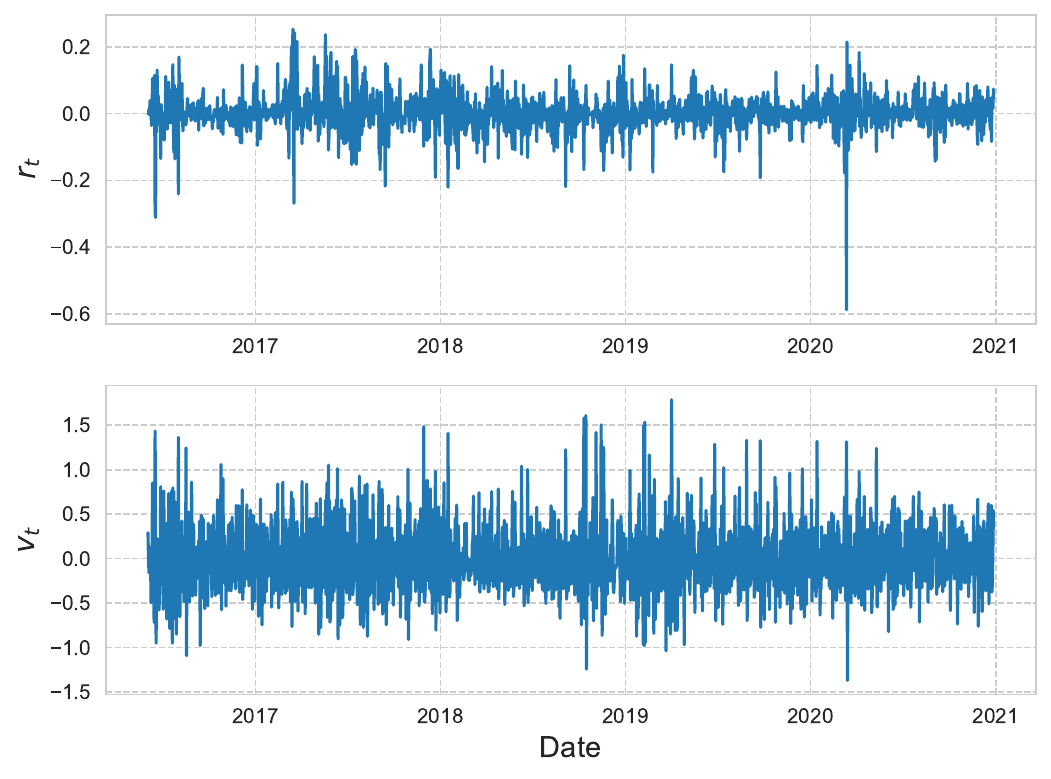}\label{fig:figETH}}
        \subfigure[XRP]{\includegraphics[width=0.45\linewidth]{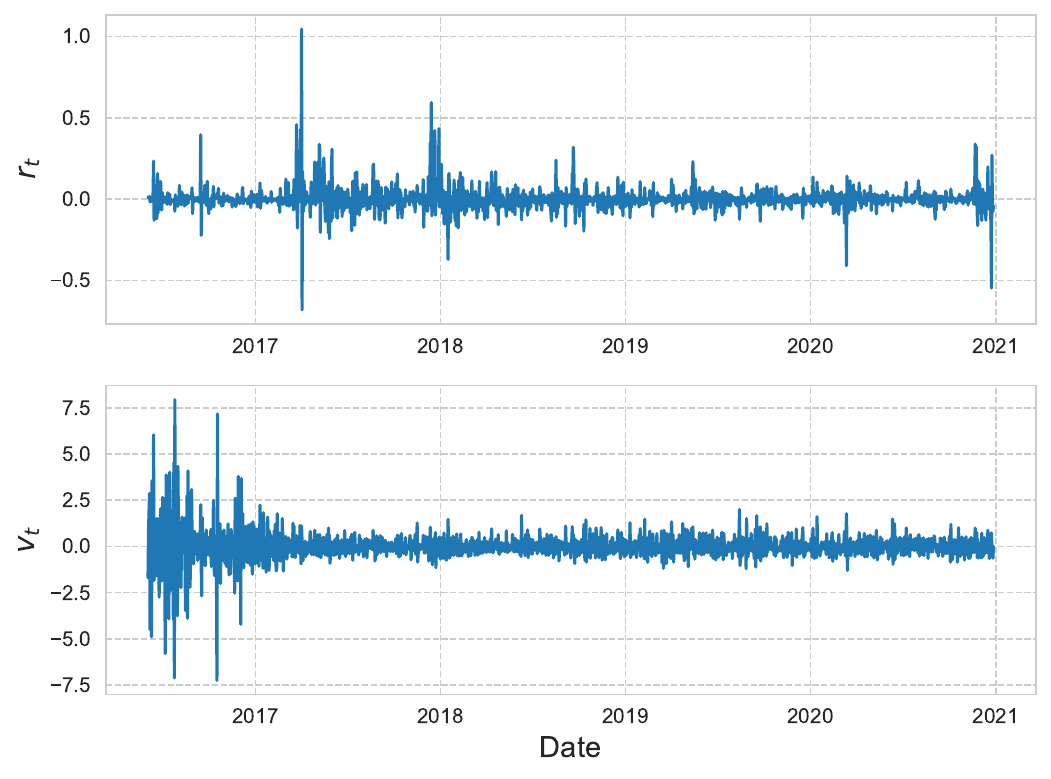}\label{fig:figXRP}}
        \subfigure[LTC]{\includegraphics[width=0.45\linewidth]{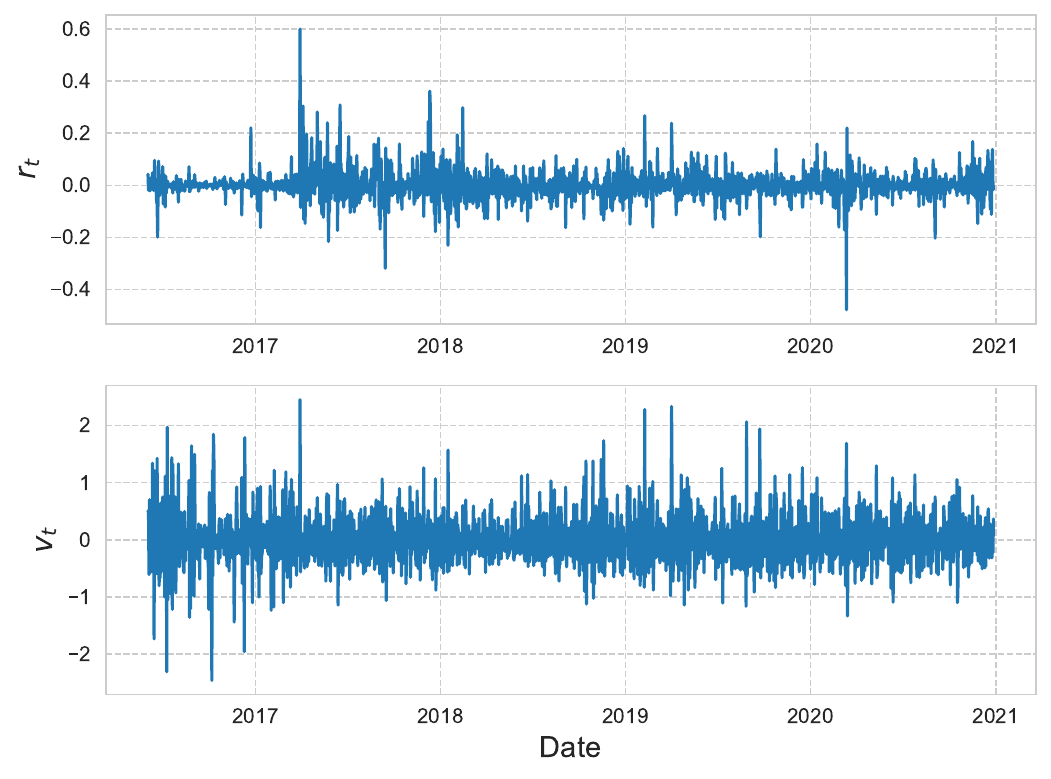}\label{fig:figLTC}}
    \end{center}
    \caption{The series of daily returns $r_t$ and the series of volatility changes $v_t$ calculated from 5 minute intervals for (a) BTC, (b) ETH, (c) XRP, and (d) LTC.}
    \label{fig:return_vola}
\end{figure*}

As a proxy of the volatility series, we employ the realized bipower variation~\cite{BarndorffNielsen2004}. Realized bipower variation is determined using high frequency price data:
\begin{align}
    \label{eq:BPV}
    \mathrm{BPV}_t = \sum_{j}|r_{t,t_j}||r_{t,t_{j+1}}|,
\end{align}
where $r_{t,t_j}$ is the return or the log-difference of price calculated from $\delta t$-minute sampling intervals, and $t_j = j\delta t$ denotes the time on day $t$. Although it is well-known that realized volatility (realized variance), the sum of squared intraday returns, is expected to converge to the integrated volatility $\sigma^2_t$ in the limit $\delta t \rightarrow 0$, its pure form is not an appropriate measure when frequent jumps and microstructure noise are observed in the series. The realized bipower variation is somewhat robust to jumps providing a model-free and consistent alternative to realized variance~\cite{BarndorffNielsen2004}. In this study, we use the 5-minute sampling interval because it avoids strong bias derived from extremely high frequencies and maintains an accurate measure of volatility~\cite{bandi2006separating, liu2015does}. The price increments (returns) and volatility increments (volatility changes) are respectively calculated as
\begin{align}
    \label{eq:price-change}
    r_t &= \ln p_{t} - \ln p_{t-1}, \\
    \label{eq:vola-change}
    v_t &= \ln \hat{\sigma}_t - \ln \hat{\sigma}_{t-1},
\end{align}
where $\hat{\sigma}_t = \sqrt{\mathrm{BPV}_t}$, and $p_t$ denotes daily closing price. To check whether the log-volatility series ($\ln \hat{\sigma}_t$) are non-stationary or not, we implemented two statistical tests; the Augmented Dickey-Fuller (ADF) test~\cite{Dickey1979} and Kwiatkowski-Philips-Schmidt-Shin (KPSS) test~\cite{Kwiatkowski1992}. We confirm that for all cases, the series are non-stationary and thus the log-differences should be considered. In particular, the ADF tests show that the null hypothesis of the existence of unit root cannot be rejected at 1\% significance level, and the KPSS tests show that the null hypothesis of sationarity is rejected at 1\% significance level (see Appendix). The data length of $r_t$ and $v_t$ equally consists of $N=1669$ for all the series we use in the analysis. The series are shown in Fig.~\ref{fig:return_vola}.

\begin{table}[ht]
\caption{\label{tab:discription_crypto_price}
Descriptive statistics for cryptocurrency market time series (returns)
}
\begin{ruledtabular}
\begin{tabular}{lrrrr}
 & BTC & ETH & XRP & LTC \\
\hline
Mean(\%) & 0.233 & 0.234 & 0.233 & 0.198 \\
Median(\%) & 0.243 & 0.042 & -0.152 & -0.067 \\
Std. Dev.(\%) & 4.174 & 5.701 & 7.397 & 5.934 \\
Max.(\%) & 23.814 & 25.274 & 104.605 & 60.051 \\
Min.(\%) & -50.435 & -58.697 & -68.039 & -47.796 \\
Skewness & -1.107 & -0.704 & 2.182 & 0.877 \\
Kurtosis & 15.557 & 9.800 & 37.335 & 12.740 \\
Jarque-Bera\footnotemark[1] & $17172^{***}$ & $6816.1^{***}$ & $98255^{***}$ & $11501^{***}$ \\
\end{tabular}
\end{ruledtabular}
\footnotetext[1]{*** denotes statistical significance at 1\% level.}
\end{table}
\begin{table}[ht]
\caption{\label{tab:discription_crypto_vola}
Descriptive statistics for cryptocurrency market time series (bipower volatility change)
}
\begin{ruledtabular}
\begin{tabular}{lrrrr}
 & BTC & ETH & XRP & LTC \\
\hline
Mean(\%) & 0.044 & 0.050 & 0.054 & 0.137 \\
Median(\%) & -2.702 & -3.017 & -4.498 & -3.454 \\
Std. Dev.(\%) & 37.672 & 39.560 & 80.847 & 47.053 \\
Max.(\%) & 204.646 & 178.746 & 794.343 & 245.149 \\
Min.(\%) & -149.576 & -136.984 & -725.121 & -246.325 \\
Skewness & 0.747 & 0.600 & 0.264 & 0.506 \\
Kurtosis & 2.446 & 1.360 & 25.454 & 2.993 \\
Jarque-Bera\footnotemark[1] & $571.44^{***}$ & $228.82^{***}$ & $44588^{***}$ & $693.64^{***}$ \\
\end{tabular}
\end{ruledtabular}
\footnotetext[1]{*** denotes statistical significance at 1\% level.}
\end{table}

In Table~\ref{tab:discription_crypto_price}, we report the descriptive statistics for the returns of the investigated cryptocurrencies. As we can see from the results of the Jarque-Bera test, all of them are remarkably far from the Gaussian distribution with high values of kurtosis and a certain degree of skewness. Similarly, descriptive statistics for the volatility changes in Table~\ref{tab:discription_crypto_vola} tell us that the series also have non-gaussian behavior, but tend to have higher variance and lower kurtosis in comparison with the return series.

\section{Methodology}
\subsection{Multifractal asymmetric detrended cross-correlation analysis}
This subsection presents a slightly modified version of the MF-ADCCA method of Cao et al.~(2014)~\cite{cao2014detrended}, where the asymmetric proxy is not directly the return series but the price index, similar to the index-based MFDFA proposed by Lee et al.~(2017)~\cite{lee2017asymmetric}. This method describes the asymmetric cross-correlations between two time series $\{x_t:t=1,\ldots,N\}$ and $\{y_t:t=1,\ldots,N\}$ in terms of whether the aggregated index shows a positive increment or a negative increment.

First, we start by constructing the profiles from the series
\begin{align*}
X(k)&= \sum_{t=1}^k(x_t-\bar{x}),\;\;t=1,\ldots,N, \\
Y(k)&= \sum_{t=1}^k(y_t-\bar{y}),\;\;t=1,\ldots,N
\end{align*}
where $\bar{x}$ and $\bar{y}$ is the average over the entire return series, respectively. We also calculate the index proxy series $I(k) = I(k-1) \exp(x_k)$ for $k=1,\ldots,N$ with $I(0)=1$, used for judging the positive and negative directions of the index series afterwards. Next, the profiles $X(k)$, $Y(k)$, and the index proxy $I(k)$ are divided into $N_s = \lfloor N/s \rfloor$ non-overlapping segments of length $s$. The division is repeated starting from the other end of the series to consider the entire profile, since $N$ is unlikely to be a multiple of $s$ and there may be remains in the profile. Thus, we have $2N_s$ segments in total for each series.

We next move on to the procedure of detrending the series. For each segment $v=1,\ldots,2N_s$ of length $s$, the local trend of the profiles are calculated by fitting a least-square degree-2 polynomial $\tilde X_v$ and $\tilde Y_v$, which is used to detrend $X(k)$ and $Y(k)$, respectively. At the same time we determine the local asymmetric direction of the index series by estimating the least-square linear fit $\tilde{I}_v(i) = a_{I_v} + b_{I_v} i\; (i=1,\ldots,s)$ for each segment. Positive (upward) or negative (downward) trends depend on the sign of the slope $b_{I_v}$.

Then the detrended covariance for each of the $2N_s$ segments is calculated as:
\begin{align*}
    f^2(s, v) =& \frac{1}{s} \sum_{i=1}^s \left | X((v-1) s+i)-\tilde{X}_v(i) \right | \nonumber \\
    & \cdot \left | Y((v-1) s+i)-\tilde{Y}_v(i) \right |
\end{align*}
for $v=1,\ldots,N_s$ and
\begin{align*}
    f^2(s, v) =& \frac{1}{s} \sum_{i=1}^s \left | X(N-(v-N_s)s+i)-\tilde{X}_v(i) \right | \nonumber \\
    & \cdot \left | Y(N-(v-N_s)s+i)-\tilde{Y}_v(i) \right |
\end{align*}
for $v=N_s+1,\ldots,2N_s$.

The upward and downward $q$-th order fluctuation functions are calculated by taking the average over all segments as:
\begin{align}
    \label{qfluctuationfunction+-}
    \nonumber
    F^+_q(s)&=\left\{ \frac{1}{M^+} \sum_{v=1}^{2N_s} \frac{1+\mathrm{sgn}(b_{I_v})}{2} \left[ f^2(s,v) \right]^{q/2}\right\}^{1/q},\\
    F^-_q(s)&=\left\{ \frac{1}{M^-} \sum_{v=1}^{2N_s} \frac{1-\mathrm{sgn}(b_{I_v})}{2} \left[ f^2(s,v) \right]^{q/2}\right\}^{1/q},
\end{align}
for any real value $q\neq0$, and
\begin{align}
    \nonumber
    \label{qfluctuationfunction+-_q0}
    F^+_0(s)&=\exp \left\{ \frac{1}{2M^+} \sum_{v=1}^{2N_s} \frac{1+\mathrm{sgn}(b_{I_v})}{2} \ln \left[ f^2(s,v) \right]\right\},\\
    F^-_0(s)&=\exp \left\{ \frac{1}{2M^-} \sum_{v=1}^{2N_s} \frac{1-\mathrm{sgn}(b_{I_v})}{2} \ln \left[ f^2(s,v) \right]\right\},
\end{align}
for $q=0$.
$M^+ = \sum_{v=1}^{2N_s}\frac{1+\mathrm{sgn}(b_{I_v})}{2}$ and $M^- = \sum_{v=1}^{2N_s}\frac{1-\mathrm{sgn}(b_{I_v})}{2}$ respectively represent the numbers of segments with positive and negative trends under the assumption of $b_{I_v}\neq0$ for all $v=1,\ldots,2N_s$, such that $M^++M^-=2N_s$.
The $q$-th order fluctuation functions for the overall trend corresponds to the MF-DCCA method shown as:
\begin{align}
    \label{qfluctuationfunction}
    F_q(s)=\left\{ \frac{1}{2N_s} \sum_{v=1}^{2N_s} \left[ f^2(s,v) \right]^{q/2}\right\}^{1/q},
\end{align}
for $q\neq0$ and when $q=0$,
\begin{align}
    \label{qfluctuationfunction_q0}
    F_0(s)=\exp \left\{ \frac{1}{4N_s} \sum_{v=1}^{2N_s} \ln \left[ f^2(s,v) \right]\right\}.
\end{align}

If the series $x_k$ and $y_k$ are long-range power-law cross-correlated, then the $q$-th order fluctuation functions follow a power-law of the forms $F^+_q(s) \sim s^{h^+_{xy}(q)}$, $F^-_q(s) \sim s^{h^-_{xy}(q)}$, and $F_q(s) \sim s^{h_{xy}(q)}$. The long-range power-law correlation properties are represented in terms of the scaling exponent also known as the generalized Hurst exponent.

The scaling exponent can easily be calculated by performing a log-log linear regression. However, the performance of the regression more or less depends on the choice of which range of scales to be implemented. As recommended in Thompson and Wilson~(2016)~\cite{thompson2016multifractal}, we employ the scale ranging from $s_\mathrm{min} = \max(20, N/100)$ to $s_\mathrm{max} = \min(20s_\mathrm{min}, N/10)$ and using $100$ points in the regression, in order to avoid biases and maintain the validity of the estimation.

In cases of no cross-correlations, $h_{xy}(q)=0.5$ satisfies. If $h_{xy}(q)>0.5$, the cross-correlations between the series are persistent with long-memory. On the contrary, $h_{xy}(q)<0.5$ indicates that the cross-correlations between the series are anti-persistent with short-memory. The same explanation certainly holds for $h^+_{xy}(q)$ and $h^-_{xy}(q)$, but the scaling exponents of cross-correlations are individually measured for positive and negative increments.

The order $q$ implies to what degree the various magnitudes of fluctuations are to be evaluated. Scaling exponents for $q>0$, where the fluctuation function $F_q(s)$ is dominated by large fluctuations, reflect the behavior of larger fluctuations. Scaling exponents for $q<0$ reflect the behavior of smaller fluctuations since small fluctuations dominate the fluctuation function. If $h_{xy}(q)$ is independent of $q$, then the cross-correlation of the series is monofractal since the scaling behavior of the detrended covariance $F^2(s,v)$ is identical for all segments. On the other hand, if the value differs depending on $q$, small and large behaviors have different scaling properties and the cross-correlations of the series are multifractal. It should be mentioned that when $q=2$, $h_{xy}(q)$ corresponds to the Hurst exponent.

The features of the multifractality can be further explored by the R\'enyi's exponent given as
\begin{align}
    \tau_{xy}(q) = qh_{xy}(q)-1.
\end{align}
If $\tau_{xy}(q)$ is a linear function of $q$, the cross-correlation of the series is monofractal but otherwise, it is multifractal.
From the Legendre transform, the singularity spectrum is obtained as follows:
\begin{align}
    \nonumber
    \alpha &= h_{xy}(q) + qh'_{xy}(q)\\
    f_{xy}(\alpha) &= q(\alpha - h_{xy}(q))+1,
\end{align}
where $\alpha$ is the singularity of the bivariate series. The singularity spectrum width $\Delta \alpha = \alpha_{\max}-\alpha_{\min}$ represents the degree of multifractality of the bivariate series, where $\alpha_{\max}$ and $\alpha_{\min}$ are respectively the $\alpha$ values at the maximum and minimum of $f_{xy}(\alpha)$ support. In the case of monofractality, $\Delta \alpha$ heads to zero, and thus the singularity spectra is theoretically just a point. The discussions above can also be expanded to examine the multifractal properties for asymmetric cases of generalized Hurst exponents $h^+_{xy}(q)$ and $h^-_{xy}(q)$. Thus the asymmetric cases of the R\'enyi's exponent, $\tau^+_{xy}(q)$ and $\tau^-_{xy}(q)$, and the singularity spectra, $f^+_{xy}(\alpha)$ and $f^-_{xy}(\alpha)$, can be calculated as well.

It must be noticed that when the series ${x_j}$ and ${y_j}$ are identical, the method aims to study the properties of autocorrelations and the method is consistent with the A-MFDFA.

\subsection{Asymmetric DCCA coefficient}
The cross-correlation between two series with a high degree of non-stationarity and self-similarity can be quantified via the DCCA coefficient, which utilizes the analysis based on the DCCA. An asymmetric extension of the cross-correlation coefficient is proposed by Cao et al.~(2018)~\cite{cao2018multifractal}, where the coefficient of the bivariate series for the cases of when prices increase and decrease can be examined separately.

The cross-correlation coefficient of Zebende~(2011)~\cite{zebende2011dcca} and Podobonik et al.~(2011)~\cite{podobnik2011statistical} between the series $\{x_t: t=1,\ldots,N\}$ and $\{x_t: t=1,\ldots,N\}$ is derived relying on the fluctuation functions calculated from overlapping $N-s$ segments of length $s+1$ as
\begin{align}
    F^2_{\mathrm{DCCA}}(s) = \frac{1}{N-s}\sum_{i=1}^{N-s} f^2_{\mathrm{DCCA}}(s, i),
\end{align}
where $f^2_{\mathrm{DCCA}}(s, i)$ is the detrended covariance of the residuals for each segment defined as
\begin{align}
    \label{dccacoeff_detrended_covariance}
    f^2_{\mathrm{DCCA}}(s, i) = \frac{1}{s+1}\sum_{k=i}^{1+s} (R_x(k)-\tilde{R}_x(k))(R_y(k)-\tilde{R}_y(k)),
\end{align}
where $R_x(k)=\sum_{t=1}^k x_t$, $R_y(k)=\sum_{t=1}^k y_t$, and the degree-2 polynomial fits $\tilde{R}_x(k)$ and $\tilde{R}_y(k)$ are used to detrend $R_x(k)$ and $R_y(k)$. When we are considering the correlations between two identical series, the fluctuation function $F^2_{\mathrm{DCCA}}(s)$ is reduced to $F^2_{\mathrm{DFA}}(s)$ defined in terms of the DFA method.

Based on the fluctuation functions above, the cross-correlation coefficient is defined as follows:
\begin{align}
    \rho_{\mathrm{DCCA}}(s) = \frac{F^2_{\mathrm{DCCA}}(s)}{F_{\mathrm{DFA}x}(s)F_{\mathrm{DFA}y}(s)},
\end{align}
where $F^2_{\mathrm{DFA}x}(s)$ and $F^2_{\mathrm{DFA}y}(s)$ is the DFA fluctuation function for each of the series $x$ and $y$, respectively. Note that with a given scale of $s$, $F^2_{\mathrm{DCCA}}(s)$ represents cross-correlation features at that scale, and $F^2_{\mathrm{DFA}}(s)$ represents autocorrelation features of each series at scale $s$. This coefficient, therefore, reveals the degree of cross-correlations for various scales.

In the asymmetric DCCA coefficient, the cross-correlation coefficients for the upward and downward trends are taken in consideration as follows:
\begin{align}
    \nonumber
    \rho_{\mathrm{DCCA}+}(s) = \frac{F^2_{\mathrm{DCCA}+}(s)}{F_{\mathrm{DFA}x+}(s)F_{\mathrm{DFA}y+}(s)},\\
    \rho_{\mathrm{DCCA}-}(s) = \frac{F^2_{\mathrm{DCCA}-}(s)}{F_{\mathrm{DFA}x-}(s)F_{\mathrm{DFA}y-}(s)},
\end{align}
where $F^2_{\mathrm{DCCA}+}(s)$ and $F^2_{\mathrm{DCCA}-}(s)$ are obtained from the calculation process in line with equation~\eqref{qfluctuationfunction+-}, using the detrended covariance of the residuals for each segment shown in equation~\eqref{dccacoeff_detrended_covariance}. The coefficients $\rho_{\mathrm{DCCA}}$, $\rho_{\mathrm{DCCA+}}$, and $\rho_{\mathrm{DCCA-}}$ range from -1 to 1, and the value equal to 1 indicates the existence of a perfect cross-correlation, and -1 indicates perfect anti-cross-correlation.

\section{Results and discussions}
This section explores asymmetric multifractal features of price-volatility cross-correlations and quantifies their coupling levels to clarify the presence of asymmetric volatility effects in cryptocurrency markets.

\subsection{Multifractal and asymmetric properties of cross-correlations}
Before we conduct the MF-ADCCA method, we first test the presence of cross-correlations between price changes and volatility changes to confirm that application of DFA-based methods is appropriate for the analyses. We apply the statistic test proposed by Podobnik et al.~(2009)~\cite{podobnik2009quantifying} to check the presence of cross-correlations between the bivariate series. The cross-correlation statistic for the series $\{x_i\}$ and $\{y_i\}$ of equal length $N$ is defined as:
\begin{align}
    \label{eq:Qcc}
    Q_{cc}(m) = N^2\sum_{i=1}^m \frac{X^2_i}{N-i},
\end{align}
where $X_i$ is the cross-correlation function defined as:
\begin{align}
    \label{eq:Qcc2}
    X_i = \frac{\sum_{k=i+1}^N x_k y_{k-i}}{\sqrt{\sum_{k=1}^N x^2_k \sum_{k=1}^N y^2_k}}.
\end{align}
Since the statistic $Q_{cc}(m)$ is approximately $\chi^2(m)$ distributed with $m$ degrees of freedom, it can be used to test the null hypothesis that the first $m$ cross-correlation coefficients are nonzero. If the value of $Q_{cc}(m)$ is larger than the critical value of $\chi^2(m)$, the null hypothesis is rejected and thus the series have a significant cross-correlation.

\begin{figure}[htbp]
    \begin{center}
    \includegraphics[width=\linewidth]{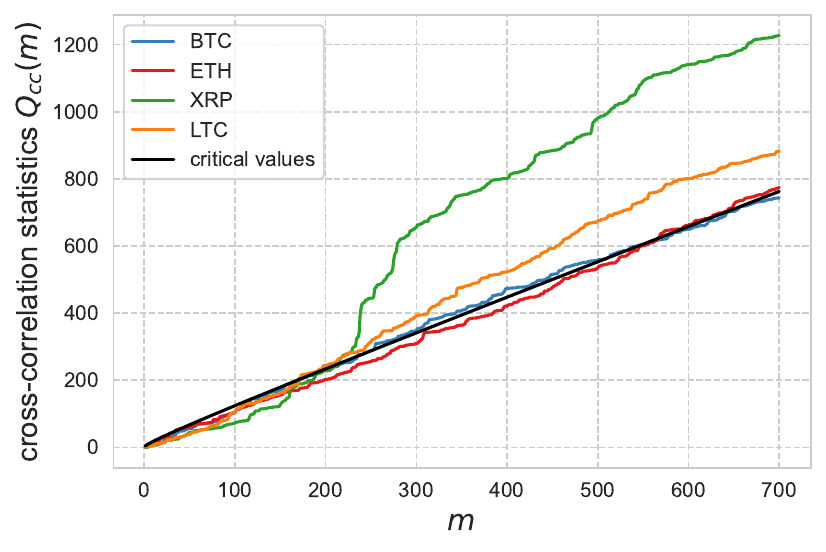}
    \caption{\label{fig:Qcc}Cross-correlation statistics between the return series and volatility change series of the four major cryptocurrencies. The black line represents the critical values at the 5\% level of significance.}
\end{center}
\end{figure}

The cross-correlation test statistics in eqs.\eqref{eq:Qcc} and~\eqref{eq:Qcc2} for price changes and volatility changes of the four cryptocurrencies are calculated with various degrees of freedom $m$, ranging from 1 to 700. The results are shown in Fig.~\ref{fig:Qcc} together with the critical values of the $\chi^2(m)$ distribution at the 5\% level of significance. We find that for all the investigated cryptocurrencies, the statistic $Q_{cc}(m)$ deviates from the corresponding critical value, indicating that there are nonlinear cross-correlations between price changes and volatility changes. For XRP and LTC, the test statistic deviates from the critical value more than those of BTC and ETH. This implies the presence of stronger nonlinear cross-correlations in the minor cryptocurrencies compared to the major ones.

\begin{figure*}[htbp]
    \begin{center}
        \subfigure[BTC]{\includegraphics[width=0.9\linewidth]{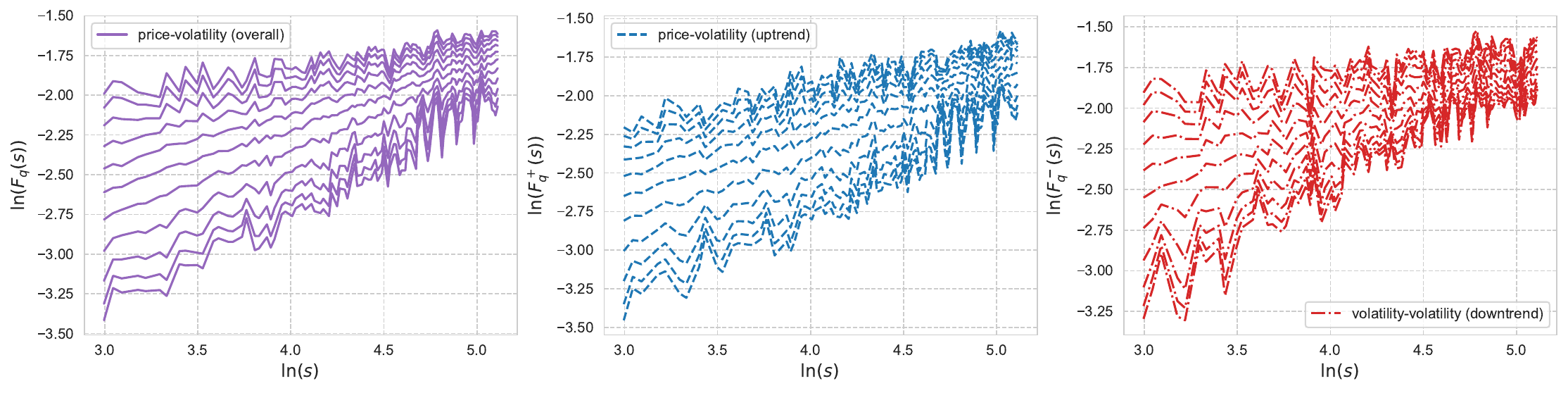}\label{fig:FqsBTC}}
        \subfigure[ETH]{\includegraphics[width=0.9\linewidth]{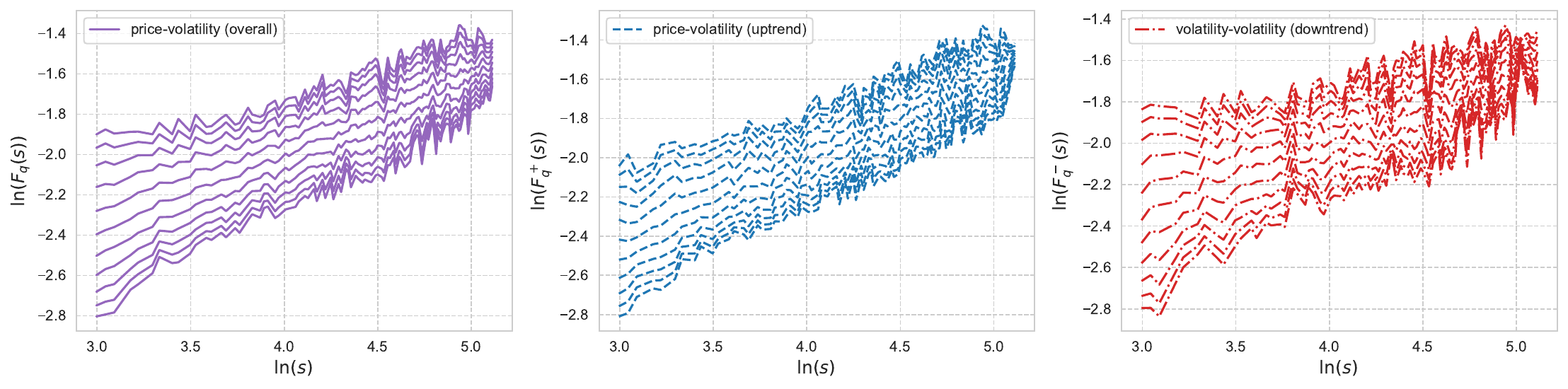}\label{fig:FqsETH}}
        \subfigure[XRP]{\includegraphics[width=0.9\linewidth]{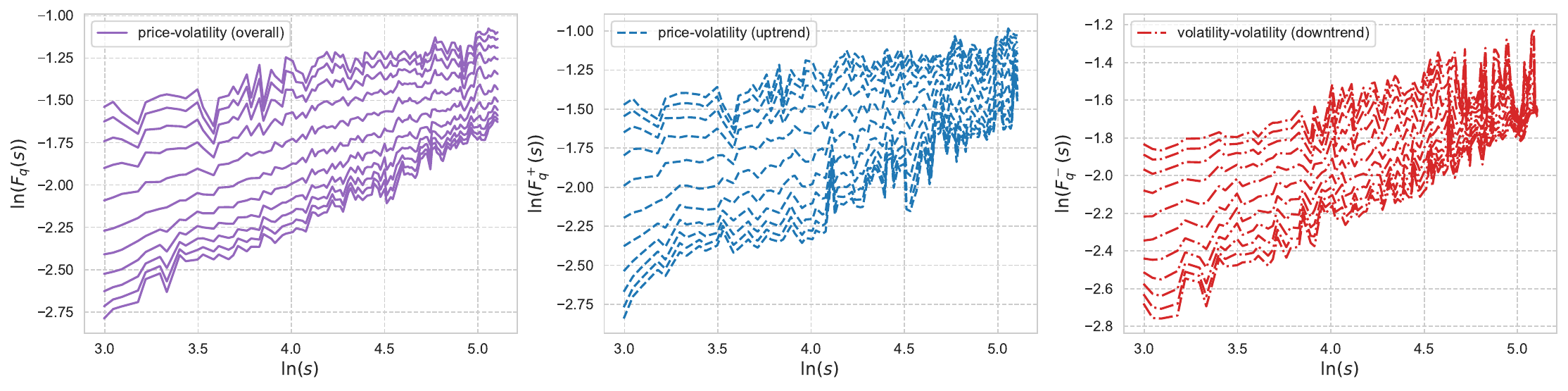}\label{fig:FqsXRP}}
        \subfigure[LTC]{\includegraphics[width=0.9\linewidth]{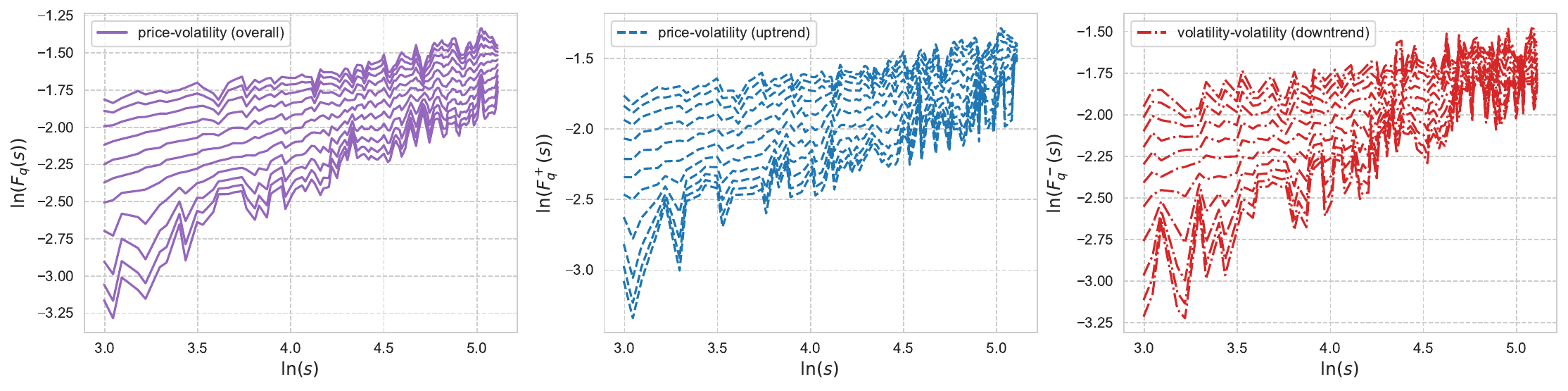}\label{fig:FqsLTC}}
    \end{center}
    \caption{Log-log plots of $F_q(s), F^+_q(s)$ and $F^-_q(s)$ versus time scale $s$ for the four cryptocurrency series of (a) BTC, (b) ETH, (c) XRP, and (d) LTC. We show the cases of $q=-10, -8, -6, \ldots, 10$. The power-law relations indicate that the bivariate series have long-range power-law cross-correlations.}
    \label{fig:Fqs}
\end{figure*}

Now that we have verified the existence of nonlinear cross-correlations in the bivariate series, we next analyze the multifractal properties of asymmetric cross-correlations for each cryptocurrency via the MF-ADCCA method. Figure~\ref{fig:Fqs} shows the $q$-th order fluctuation functions calculated from the returns and volatility changes with various $q$ ranging from -10 to 10. The fluctuation functions under different situations of bull and bear markets (uptrend and downtrend) are also depicted with the overall trend. For all cases, we observe that the fluctuation functions generally follow a power-law against the scale, which means that the cross-correlations between the bivariate series have a long-range power-law property. Therefore the MF-ADCCA is expected to be an effective method for analyzing cross-correlations, along with the asymmetry between uptrend and downtrend cross-correlations.

\begin{figure*}[thbp]
    \begin{center}
        \subfigure[BTC]{\includegraphics[width=0.85\columnwidth]{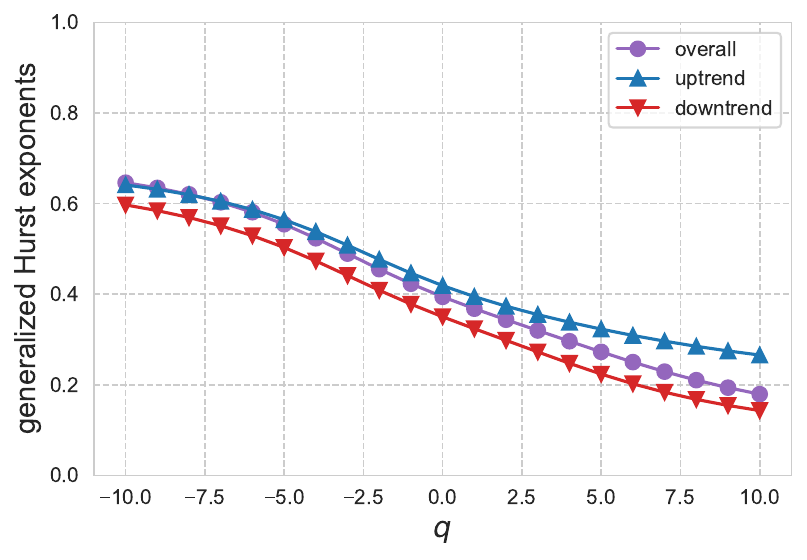}\label{fig:ghurstBTC}}
        \subfigure[ETH]{\includegraphics[width=0.85\columnwidth]{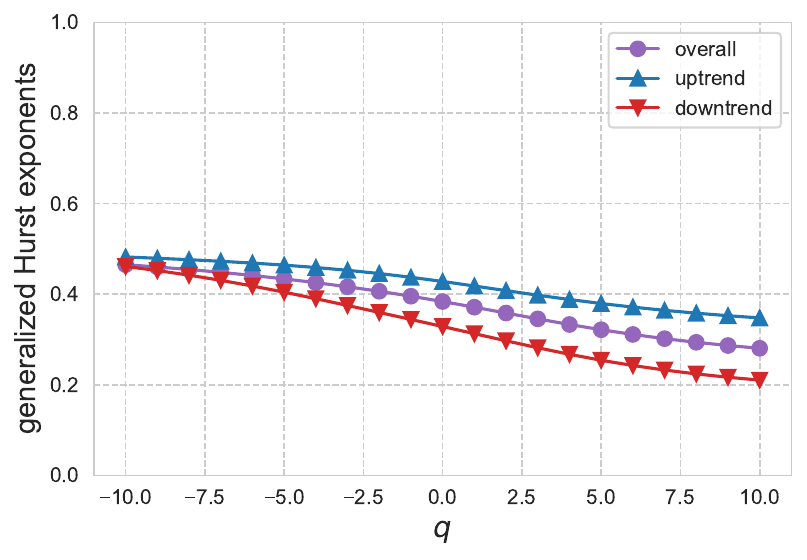}\label{fig:ghurstETH}}
        \subfigure[XRP]{\includegraphics[width=0.85\columnwidth]{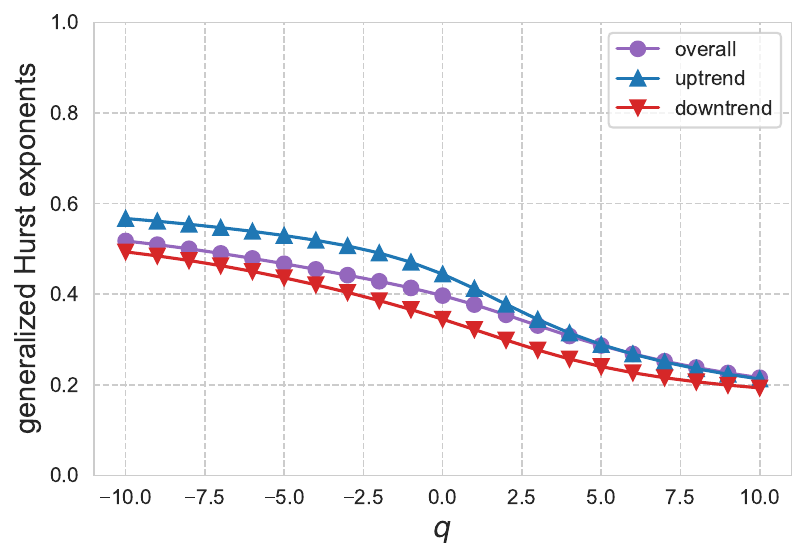}\label{fig:ghurstXRP}}
        \subfigure[LTC]{\includegraphics[width=0.85\columnwidth]{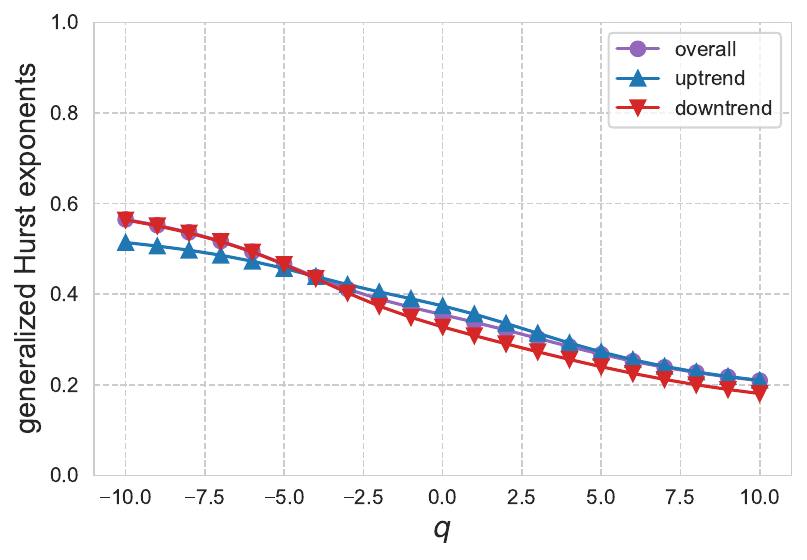}\label{fig:ghurstLTC}}
    \end{center}
    \caption{Relationship between the generalized Hurst exponents and the order $q$ for the cases of (a) BTC, (b) ETH, (c) XRP, and (d) LTC.}
    \label{fig:ghurst}
\end{figure*}

Figure~\ref{fig:Fqs} shows that the behavior of power-law cross-correlations varies among market situations of different trends. To measure the degree of asymmetry of the cross-correlations, we calculate the metric defined as
\begin{align}
    \Delta h_{xy}(q) = h^+_{xy}(q) - h^-_{xy}(q),
\end{align}
for given $q$. The greater the value, the greater the asymmetric behavior in terms of different trends. If $\Delta h_{xy}(q) > 0$ $(\Delta h_{xy}(q) < 0)$, the cross-correlation for uptrend situations has a larger (smaller) exponent compared to those of downtrend situations. When the bivariate series are essentially identified by the same scaling exponent, $\Delta h_{xy}(q)$ is theoretically zero and the two series have symmetric cross-correlations. We calculate the cases of $q=-10$ (small fluctuations), $q=2$ (corresponding to the Hurst exponent), and $q=10$ (large fluctuations). We clearly find in Table~\ref{tab:degree_crypto} that regardless of small and large fluctuations, $\Delta h_{xy}(q)$ is positive for all the investigated cryptocurrencies (except for the case of $q=-10$ in LTC). Fig.~\ref{fig:ghurst} also supports these findings where $h^+_{xy}(q)$ is larger than $h^-_{xy}(q)$. Cross-correlations of price-volatility in the uptrend markets generally have slightly higher persistency at all levels of fluctuations compared to those in the downtrend markets. To investigate the statistical validity of the asymmetric multifractal degree, we implement Monte Carlo simulations and obtain confidence intervals of $\Delta h_{xy}(q)$ calculated from 1000 generated series of returns and realized bipower volatility changes. The daily returns and realized bipower volatility changes are constructed based on the shuffled series of the original 5-minute high frequency returns. We find that BTC, ETH, and XRP exhibit asymmetry at the 1\% or 5\% significance level, while LTC have rather insignificant asymmetry. We also find that the major currencies of BTC and ETH present significant asymmetry especially in large fluctuations, but the minor currencies tend to present more asymmetry in smaller fluctuations.

\begin{table}[tbh]
\caption{\label{tab:degree_crypto}
Asymmetric degree of price-volatility cross-correlations in terms of uptrend and downtrend price movements shown together with the multifractal degree for each of the four cryptocurrencies. Note that ***, **, and * denote 1\%, 5\%, and 10\% significance levels, respectively.}
\begin{ruledtabular}
\begin{tabular}{lllll}
 & $\Delta h_{xy}(-10)$ & $\Delta h_{xy}(2)$ & $\Delta  h_{xy}(10)$ & $D_{xy}$ \\
\hline
BTC & $0.044$ & $0.075^{**}$ & $0.123^{***}$ & $0.233$ \\
ETH & $0.021$ & $0.111^{***}$ & $0.158^{***}$ & $0.127$ \\
XRP & $0.074^{**}$ & $0.079^{**}$ & $0.020$ & $0.151$ \\
LTC & $-0.050$ & $0.045$ & $0.029$ & $0.178$ \\
\end{tabular}
\end{ruledtabular}
\end{table}

The presence of multifractality can be examined by looking into whether or not the generalized Hurst exponents are dependent on its order $q$. Given that $h_{xy}(q)$ decreases as $q$ increases in Fig.~\ref{fig:ghurst}, $h_{xy}(q)$ is not constant for $q$ and hence multifractal behaviors exsist in the cross-correlations between the bivariate series. To numerically explain the deviation from monofractality and efficiency, we use the market efficiency measure (MDM) of Wang et al.~(2009)~\cite{wang2009analysis} defined as
\begin{align}
    D_{xy}=\frac{1}{2}(|h_{xy}(-10)-0.5|+|h_{xy}(10)-0.5|).
\end{align}
If $D_{xy}$ is zero or close to zero, then the relationship between the series are efficient. Larger values of $D_{xy}$ indicate higher inefficiency, and smaller values indicate lower inefficiency. This metric is useful to determine the ranking of the (in)efficiency degree~\cite{mensi2017global}. From the results shown in Table~\ref{tab:degree_crypto}, we suggest that BTC is the most inefficient with $D_{xy}=0.233$, and ETH is the most closest to efficiency with $D_{xy}=0.127$.

\begin{figure*}[htbp]
    \begin{center}
        \subfigure[BTC]{\includegraphics[width=0.85\columnwidth]{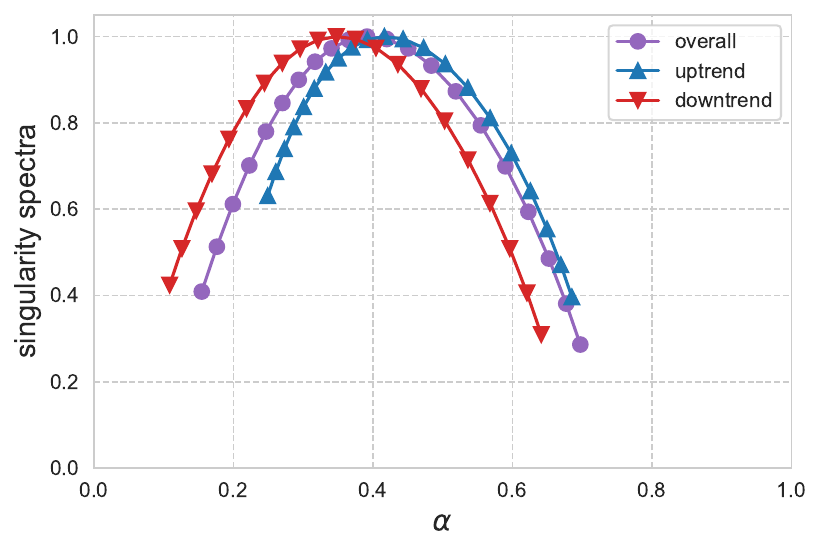}\label{fig:faBTC}}
        \subfigure[ETH]{\includegraphics[width=0.85\columnwidth]{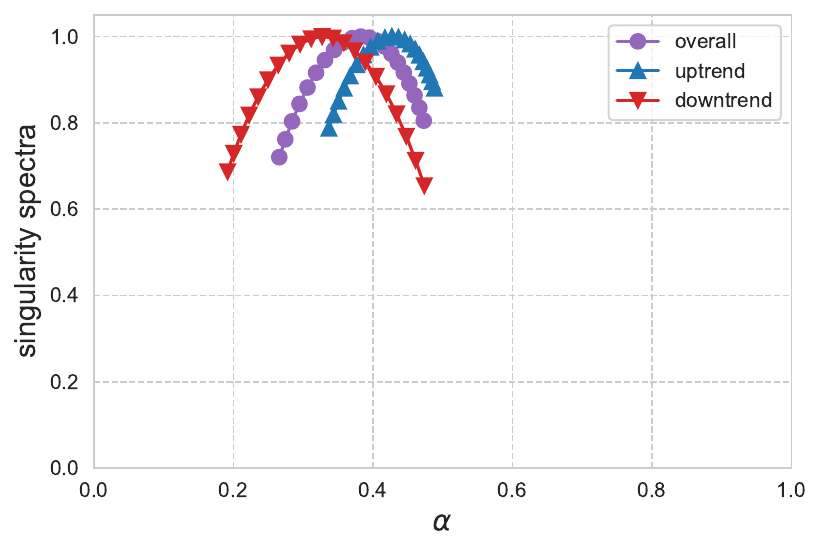}\label{fig:faETH}}
        \subfigure[XRP]{\includegraphics[width=0.85\columnwidth]{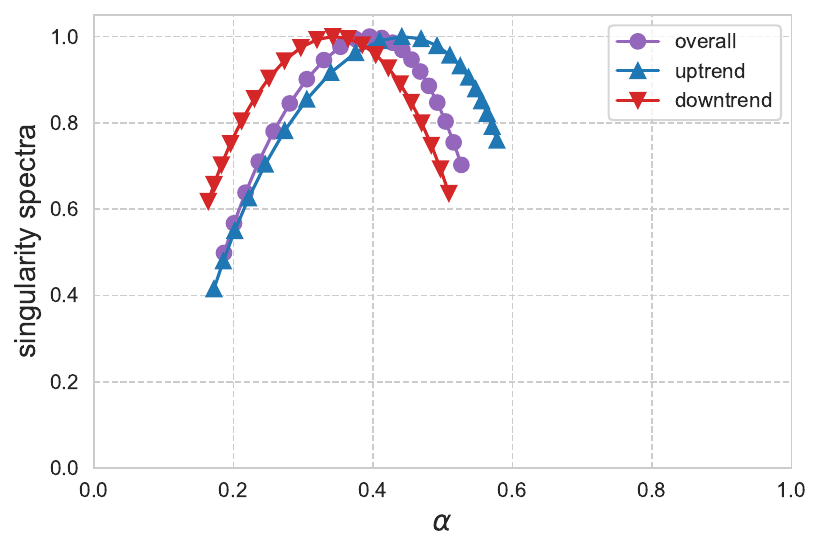}\label{fig:faXRP}}
        \subfigure[LTC]{\includegraphics[width=0.85\columnwidth]{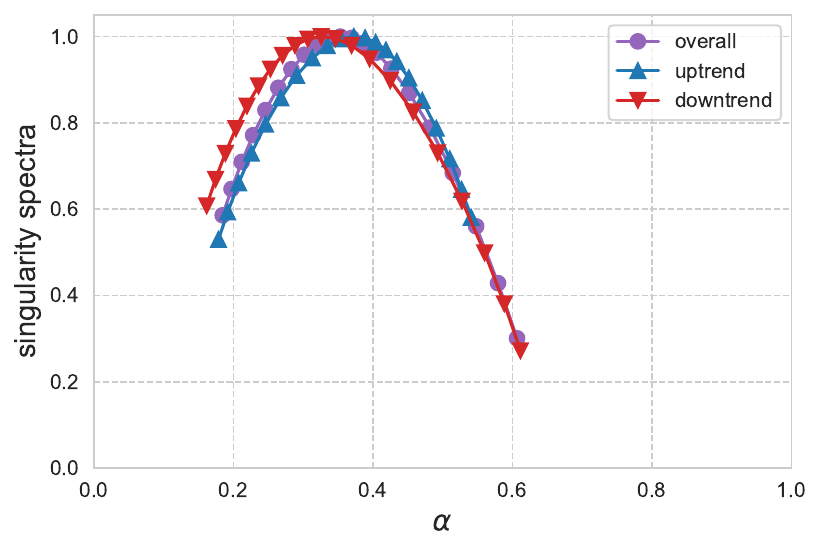}\label{fig:faLTC}}
    \end{center}
    \caption{Singularity spectra $f_{xy}(\alpha)$, $f^+_{xy}(\alpha)$, and $f^-_{xy}(\alpha)$ for the cases of (a) BTC, (b) ETH, (c) XRP, and (d) LTC.}
    \label{fig:fa}
\end{figure*}

To further study the multifractal properties, we display the singularity spectra $f_{xy}(\alpha)$, $f^+_{xy}(\alpha)$, and $f^-_{xy}(\alpha)$ in Fig.~\ref{fig:fa} where the asymmetric market trends are considered. The spectra are quite broad as expected, and the width differs with respect to cryptocurrencies and market trends. In particular, among the investigated markets, spectrum width in BTC for the overall trends present the largest degree of multifractality of $\Delta\alpha=0.5427$. On the contrary, spectrum width in ETH for uptrends present the smallest degree of $\Delta\alpha=0.1513$. In addition, we discuss the multifractal features from another metric of the asymmetric spectrum parameter, $A_{\alpha}=\frac{\Delta\alpha_L - \Delta\alpha_R}{\Delta\alpha_L + \Delta\alpha_R}$, where $\Delta\alpha_L = \alpha_0 - \alpha_{\mathrm{min}}$, $\Delta\alpha_R = \alpha_{\mathrm{max}} - \alpha_0$, and $\alpha_0$ is the value of $\alpha$ at the maximum of the singular spectrum~\cite{drozdz2015detecting}. Note that the asymmetry here is not in terms of different market trends but the distortion of the singularity spectrum $f_{xy}(\alpha)$. The metric $A_\alpha$ provides information to identifying the compositions of the bivariate series. If $A_{\alpha} > 0$ ($A_{\alpha} < 0$), the singular spectrum has a left-sided (right-sided) asymmetry, which indicates that the scaling properties are determined by $q>0$ ($q<0$) and hence larger fluctuations (smaller fluctuations) dominate the multifractal behavior~\cite{drozdz2015detecting}. Therefore, it can be interpreted that multifractality owes to larger fluctuations for the left-sided case, and the opposite for the right-sided case. If $A_{\alpha} = 0$, the left and right sides of spectrum width are equivalent, and thus the small and large fluctuations operate equally on multifractality.

\begin{table*}[htbp]
\caption{\label{tab:spectradegree_crypto}
The asymmetric degree of singularity spectra $A_\alpha$ and $\alpha_0$, $\alpha_{\mathrm{max}}$, and $\alpha_{\mathrm{min}}$ with respect to price-volatility coupling. We show also the asymmetric spectrum parameter and the values $\alpha$ for asymmetric market trends (uptrend and downtrend).
}
\begin{ruledtabular}
\begin{tabular}{lrrrrrrrrrrrr}
 &\multicolumn{4}{l}{Overall}&\multicolumn{4}{l}{Uptrend}&\multicolumn{4}{l}{Downtrend}\\
& $A_\alpha$ & $\alpha_0$ & $\alpha_{\mathrm{max}}$ & $\alpha_{\mathrm{min}}$ & $A^+_\alpha$ & $\alpha^+_0$ & $\alpha^+_{\mathrm{max}}$ & $\alpha^+_{\mathrm{min}}$ & $A^-_\alpha$ & $\alpha^-_0$ & $\alpha^-_{\mathrm{max}}$ & $\alpha^-_{\mathrm{min}}$ \\
\hline
BTC & -0.127 & 0.392 & 0.697 & 0.155 & -0.233 & 0.416 & 0.685 & 0.249 & -0.103 & 0.348 & 0.641 & 0.109 \\
ETH & 0.129 & 0.383 & 0.473 & 0.266 & 0.193 & 0.427 & 0.488 & 0.337 & -0.039 & 0.327 & 0.474 & 0.192 \\
XRP & 0.226 & 0.395 & 0.527 & 0.187 & 0.328 & 0.441 & 0.578 & 0.172 & 0.036 & 0.343 & 0.509 & 0.164 \\
LTC & -0.200 & 0.352 & 0.607 & 0.185 & 0.071 & 0.373 & 0.541 & 0.179 & -0.270 & 0.326 & 0.612 & 0.162 \\
\end{tabular}
\end{ruledtabular}
\end{table*}

We report in Table~\ref{tab:spectradegree_crypto} the asymmetric spectrum parameter $A_\alpha$ and values of $\alpha$ for the three different market trends: overall, uptrend, and downtrend. In real-world finance data, left-sided spectrum is more common and in such case, it is reasonable to have peculiar features in the larger fluctuations and a noise-like behavior in the smaller fluctuations~\cite{drozdz2015detecting}. However, we find uncommon features in some cases. The asymmetric spectrum parameters $A_{\alpha}$, $A^+_{\alpha}$, and $A^-_{\alpha}$ take negative values in BTC and LTC markets (except for $A^+_{\alpha}$ in LTC taking a positive value), and the right-skewed spectra explain that smaller events play a more important role in the underlying multifractality. In contrast, larger events contribute more to the multifractal behavior in the XRP market because the asymmetric spectrum parameters take positive values. It is interesting to mention that regardless of the market trends, multifractality of the price-volatility coupling for BTC and XRP exhibit the same behavior of asymmetry $f_{xy}(\alpha)$ either with the right- or left-sided spectrum, i.e., the same distortion. On the other hand, ETH and LTC show different multifractal properties depending on market trends, where large fluctuations are dominant in bull markets but small fluctuations are dominant in bear markets. We find that LTC tends to show relatively stronger right-skewed $f_{xy}(\alpha)$ with smaller values of $A_\alpha$, whereas XRP shows the strongest left-skewed $f_{xy}(\alpha)$ with the largest values of $A_\alpha$ for all market trends. In addition to the right-skewed property of BTC and LTC, they tend to have wider spectrum width $\Delta \alpha$, demonstrating that these markets exhibit a highly complex price-volatility behavior.

Although the bivariate extension of scaling exponents and singularity spectra captures power-law cross-correlations and helps characterize complex behaviors between the two series, the discussion without each of its univariate exponents and spectra may not provide a satisfactory interpretation.
Focusing on the bivariate Hurst exponent $(H_{xy}=h_{xy}(2))$, many empirical studies have reported that $H_{xy}>\frac{1}{2}(H_x+H_y)$ or $H_{xy}=\frac{1}{2}(H_x+H_y)$~\cite{He2011,Owiecimka2014}, whereas numerical and theoretical studies have found that $H_{xy}<\frac{1}{2}(H_x+H_y)$ or $H_{xy}=\frac{1}{2}(H_x+H_y)$~\cite{Sela2012}. Kristoufek (2013)~\cite{Kristoufek2013} defines a mixed-correlated ARFIMA (MC-ARFIMA) process, which allows for generating power-law cross-correlations and controlling the $H_{xy}$ parameter, and shows that there is no combination of parameters leading to $H_{xy}>\frac{1}{2}(H_x+H_y)$. A theoretical basis in terms of the frequency domain using the squared spectrum coherency also supports the finding that $H_{xy}>\frac{1}{2}(H_x+H_y)$ is impossible~\cite{Kristoufek2015}.
However, the empirical estimation of $H_{xy}$ or specifically its comparison with $\frac{1}{2}(H_x+H_y)$ can lead to unreliable and controversial results due to the underlying bias such as short-term dependence bias, bias in the presence of heavy tails, and finite sample bias, which is often observed in the econophysic work of empirical studies~\cite{Kristoufek2020}. We find that among the investigated cryptocurrencies, BTC, ETH, and LTC satisfy $H_{xy}<\frac{1}{2}(H_x+H_y)$. On the other hand, XRP shows higher bivariate Hurst exponent than the average of two separate ones, with $H_{xy}=0.355$ and $\frac{1}{2}(H_x+H_y)=0.342$, implying the biased results. Towards an intuitive interpretation of cross-correlations without relying solely on $H_{xy}$, Kristoufek (2020)~\cite{Kristoufek2020} suggests two ways to overcome the issue: utilizing the DCCA-based cross-correlation coefficient and focusing on the case of $H_{xy}<\frac{1}{2}(H_x+H_y)$. We focus on the former method, since it is straightforward to implement and the DCCA measure can lead to find novel relationships between price and volatility in cryptocurrency markets. Since our discussion so far has been based on the multifractal analysis, not only the issue of Hurst exponent but also the issue of generalized Hurst exponent should be addressed. The estimation of $h_{xy}(q)$ and its comparison with $\frac{1}{2}(h_x(q)+h_y(q))$, as well as in the framework of the singular spectra $f_{xy}(\alpha)$, would play a crucial role in giving interpretation when multifractal properties are present in cross-correlations. A further work that explains the theoretical aspect of these connections is left for future work.

\subsection{Assessment of asymmetric volatility}
\begin{figure*}[htbp]
    \begin{center}
        \subfigure[BTC]{\includegraphics[width=0.85\columnwidth]{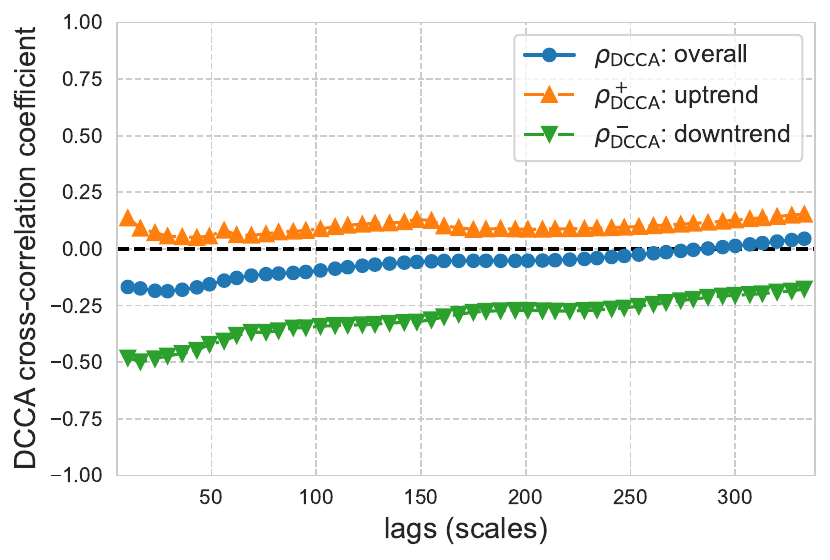}\label{fig:roBTC}}
        \subfigure[ETH]{\includegraphics[width=0.85\columnwidth]{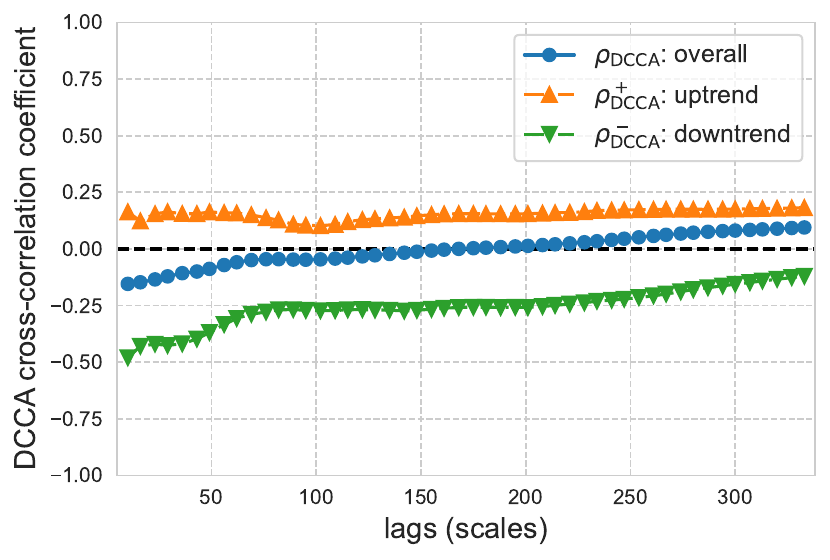}\label{fig:roETH}}
        \subfigure[XRP]{\includegraphics[width=0.85\columnwidth]{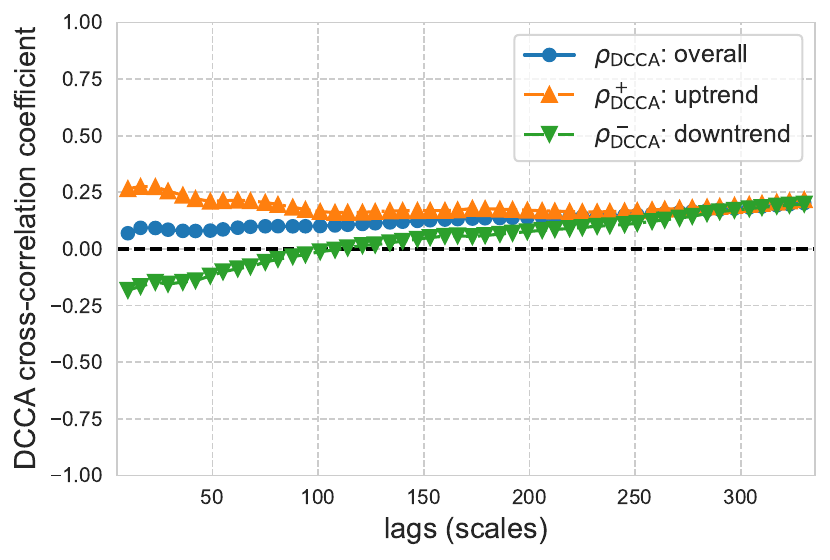}\label{fig:roXRP}}
        \subfigure[LTC]{\includegraphics[width=0.85\columnwidth]{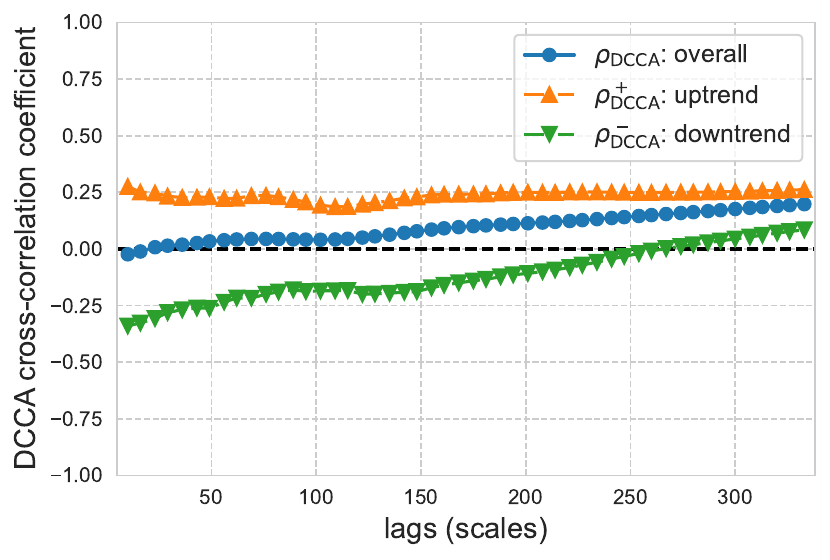}\label{fig:roLTC}}
    \end{center}
    \caption{DCCA cross-correlation coefficients $\rho_{\mathrm{DCCA}}(s)$, $\rho_{\mathrm{DCCA}+}(s)$, and $\rho_{\mathrm{DCCA}-}(s)$ between the price-volatility relationships under various scales $s$ for the cases of (a) BTC, (b) ETH, (c) XRP, and (d) LTC. The scales represent the lag of days.}
    \label{fig:ro}
\end{figure*}

We next apply the DCCA coefficient analysis to quantify the asymmetric cross-correlations between price and volatility and to examine the asymmetry volatility effects in cryptocurrency markets. Fig.~\ref{fig:ro} depicts the coefficients for the overall trend, $\rho_\mathrm{DCCA}(s)$, and the upward (bull) and downward (bear) market trends, $\rho_{\mathrm{DCCA}+}(s)$ and $\rho_{\mathrm{DCCA}-}(s)$, for various time scales $s$ ranging from 10 to 334 days. For the entire period (overall trend), the coefficients are not so far from zero at all time scales.  Analyzing the case of the overall trend appears to suggest that price and volatility are less interrelated. However, once we consider uptrend and downtrend, we can realize different pictures of the markets. The asymmetric DCCA coefficient approach enables us to separately figure out the interrelationship between price and volatility under bull and bear regimes.

As shown in Fig.~\ref{fig:ro}, the coefficients are positive for uptrend markets but negative for downtrend markets. This reconfirms that both positive and negative price changes have a certain degree of the impact on volatility. When $|\rho_{\mathrm{DCCA}-}(s)| > |\rho_{\mathrm{DCCA}+}(s)|$ is satisfied, i.e., price and volatility are more strongly cross-correlated in downtrend markets than in uptrend markets, we can say that asymmetric volatility is present at a certain time scale. On the contrary, when $|\rho_{\mathrm{DCCA}-}(s)| < |\rho_{\mathrm{DCCA}+}(s)|$ holds, we can say that inverse-asymmetric volatility is present where uptrend markets have stronger price-volatility cross-correlations.

\begin{table*}[htbp]
\caption{\label{tab:garch}
Estimate results of conditional variance for cryptocurrencies. The EGARCH model and GJR-GARCH model are estimated over the period from June 3, 2016 to December 28, 2020. Standard errors of estimates are reported in parentheses. Note that ***, **, and * denote 1\%, 5\%, and 10\% significance levels, respectively. $Q^2(10)$ is the square Q-statistic and the p-values are presented in brackets.
}
\begin{ruledtabular}
\begin{tabular}{lllllllll}
 &\multicolumn{4}{l}{EGARCH}&\multicolumn{4}{l}{GJR-GARCH} \\
 \hline
 & BTC & ETH & XRP & LTC & BTC & ETH & XRP & LTC \\
 \hline
 $\omega$ & $-0.6827^{***}$ & $-0.7089^{***}$ & $-1.0252^{***}$ & $-0.3762^{***}$ & $0.0001^{***}$ & $0.0003^{***}$ & $0.0004^{***}$ & $0.0001^{***}$ \\
 & (0.0661) & (0.0712) & (0.0422) & (0.0303) & (0.0000) & (0.0000) & (0.0000) & (0.0000) \\
 $\alpha_1$ & $0.2609^{***}$ & $0.2893^{***}$ & $0.4441^{***}$ & $0.1758^{***}$ & $0.1217^{***}$ & $0.1585^{***}$ & $0.4089^{***}$ & $0.0775^{***}$\\
 & (0.0208) & (0.0185) & (0.0182) & (0.0129) & (0.0160) & (0.0172) & (0.0269) & (0.0101) \\
 $\alpha_2$ & $\textbf{-0.0586}^{***}$ & $\textbf{-0.0071}$ & $\textbf{0.0828}^{***}$ & $\textbf{0.0193}^{***}$ & $\textbf{0.1162}^{***}$ & $\textbf{0.0176}$ & $\textbf{-0.1161}^{***}$ & $\textbf{-0.0256}^{***}$ \\
 & (0.0080) & (0.0097) & (0.0097) & (0.0060) & (0.0165) & (0.0171) & (0.0257) & (0.0082) \\
 $\beta$ & $0.9217^{***}$ & $0.9135^{***}$ & $0.8706^{***}$ & $0.9549^{***}$ & $0.7769^{***}$ & $0.7649^{***}$ & $0.6297^{***}$ & $0.8997^{***}$ \\
 & (0.0083) & (0.0106) & (0.0060) & (0.0040) & (0.0176) & (0.0195) & (0.0141) & (0.0096) \\
 \hline
 Log likelihood & 3079.33 & 2553.73 & 2412.35 & 2419.25 & 3083.58 & 2556.66 & 2428.99 & 2466.02 \\
 AIC & -3.6818 & -3.0524 & -2.8831 & -2.9775 & -3.6869 & -3.0559 & -2.9030 & -2.9473 \\
 $Q^2(10)$ & 2.4476 & 8.4555 & 4.7584 & 1.7640 & 4.2086 & 10.143 & 3.2243 & 1.2243 \\
 & [0.992] & [0.584] & [0.907] & [0.998] & [0.937] & [0.428] & [0.976] & [1.000] \\
\end{tabular}
\end{ruledtabular}
\end{table*}

The results provide clear evidence that for BTC and ETH, price and volatility are more strongly cross-correlated in downtrend markets than in uptrend markets. In particular, we find $\rho_{\mathrm{DCCA}-}(s) \approx-0.5$ and $\rho_{\mathrm{DCCA}+}(s) \approx0$ for BTC, and $\rho_{\mathrm{DCCA}-}(s) \approx-0.5$ and $\rho_{\mathrm{DCCA}+}(s) \approx0.15$ for ETH under approximately one-month time scales $(s\approx 30)$. This implies the absence of uninformed noise traders during bull markets, but they are active during bear markets. Similar results are observed when studied at longer time scales of half a year $(s\approx180)$, with the DCCA coefficients $\rho_{\mathrm{DCCA}-}(s) \approx-0.25$ and $\rho_{\mathrm{DCCA}+}(s) \approx0.1$ for both markets.  Higher levels of cross-correlations in downtrends are found across all time scales for BTC and ETH, suggesting the presence of asymmetric volatility dynamics.

In the case of XRP, however, we find inverse-asymmetry in volatility dynamics because $|\rho_{\mathrm{DCCA}+}(s)|$ is always larger than $|\rho_{\mathrm{DCCA}-}(s)|$ for all time scales. Uninformed noise traders are more dominant when the price increases than when the price decreases. More interestingly, LTC exhibits different volatility-asymmetry depending on which time scale the cross-correlation analysis was implemented at. At relatively longer time scales $(s>150)$, $|\rho_{\mathrm{DCCA}+}(s)|$ is larger than $|\rho_{\mathrm{DCCA}-}(s)|$, but at relatively shorter time scales, $|\rho_{\mathrm{DCCA}-}(s)|$ surpasses the value of $|\rho_{\mathrm{DCCA}+}(s)|$. In other words, inverse-asymmetric volatility is present at scales of $s>150$, but not at scales shorter than $s=150$. Such explanations cannot be confirmed by the conventional GARCH-class models, which generally focus on the effects of one-time price shock on volatility. The approach in this study challenges to a dynamical effects of price on volatility while accounting for the directions of price trends with various time scales. Focusing on the time scales, we find that cross-correlations for uptrend markets are generally constant; however, the loss of cross-correlations is observed for downtrend markets at longer time scales.

The asymmetric DCCA coefficient is derived from the fractal analysis and therefore reflects the scaling properties of the price-volatility nexus in a nonlinear way. However, the asymmetric behavior of the price-volatility is often assessed using the GARCH-class models which focus more on linear correlations. To make sure that our empirical findings based on the fractal framework show the same direction of asymmetric volatility with the conventional methods, we also implement two types of GARCH models that can explain the asymmetric effects on volatility. The first model is the exponential GARCH (EGARCH) written as follows~\cite{nelson1991conditional}:
\begin{align}
    \ln \sigma_t^2 = \omega + \alpha_1 \frac{|r_{t-1}|}{\sigma_{t-1}}+\alpha_2\frac{r_{t-1}}{\sigma_{t-1}}+\beta \ln \sigma_{t-1}^2,
\end{align}
with $r_t = \varepsilon_t \sigma_t$, where $\sigma_t^2$ is the conditional variance at time $t$, and $\varepsilon_t$ denotes an error term with i.i.d. standard Gaussian noise $\mathcal{N}(0,1)$.
The second model is the threshold GJR-GARCH written as follows~\cite{glosten1993relation}:
\begin{align}
    \sigma_{t}^{2}=\left\{\begin{array}{ll}\omega+\alpha_1 r_{t-1}^{2}+\beta \sigma_{t-1}^{2}, & r_{t-1} \geq 0 \\ \omega+(\alpha_1+\alpha_2) r_{t-1}^{2}+\beta \sigma_{t-1}^{2}, & r_{t-1}<0\end{array}\right.
\end{align}
where the term $\alpha_2 r_{t-1}^2$ is operated only when the market goes downwards. For convenience, both models are with one lag of the innovation $(p=1)$ and one lag of volatility $(q=1)$, since the selection of optimized lags often produces similar estimation results. The GARCH models above are represented by the parameters $\omega$, $\alpha_1$, $\alpha_2$, and $\beta$. Among these parameters, $\alpha_2$ is the responsible one that determines the asymmetric responses of volatility to market shocks. Significantly positive values of $\alpha_2$ for the EGARCH model imply that positive shocks increase volatility more than negative shocks. Note that the positive/negative direction is reversed for the GJR-GARCH model, where negative shocks increase the volatility more when $\alpha_2$ is positive. As shown in Table~\ref{tab:garch}, the values of $\alpha_2$ are negative in EGARCH and positive in GJR-GARCH for BTC and ETH, and hence the volatility is increased more by negative shocks. We find opposite results for the XRP and ETH because $\alpha_2$ are positive in EGARCH and negative in GJR-GARCH. In such cases, volatility is increased more by positive shocks. Note that for the ETH, the asymmetric parameter is insignificant and thus the asymmetric effect cannot be statistically confirmed by the implemented GARCH models.

The results of GARCH models appear to highlight that our empirical findings of the DCCA coefficient analysis on asymmetric cross-correlations are relevant to the asymmetric volatility dynamics. Both asymmetric-GARCH models and detrended cross-correlation analysis reveal consistent results of the underlying asymmetric/inverse-asymmetric volatility in cryptocurrency markets. Past studies have shown inverse-asymmetric volatility dynamics of cryptocurrency markets~\cite{baur2018asymmetric, cheikh2020asymmetric}.
Our findings conclude that unlike in the earlier periods, the volatility for the two major coins of BTC and ETH, are no longer inverse-asymmetric. One possible reason may be the less presence of uninformed noise traders and increasing participants of informed traders in the market, and the major coins head to maturity in recent years. Minor coins are immature with inverse-asymmetric volatility where speculated noise traders still dominate the market and play a significant role in raising the volatility especially during the uptrend periods. The appearance of stronger negative cross-correlation of price-volatility at shorter scales for LTC could be a key clue to explain the process that minor coins are expected to head for mature markets. As asymmetric cross-correlations are available at different scales, our findings can be applied to form models for optimal portfolio diversification under different investment horizons, which helps address the nonlinear interactions and asymmetric responses prevalent in the cryptocurrency market. In particular, investors can introduce a correlation matrix based on the asymmetric DCCA coefficient as an alternative to the conventional one, allowing investment allocations to depend on time scales. The performance of such portfolios and whether the diversification strategy is effective for making profit are left for future works.

\section{Conclusion}
This paper examined the nexus between daily price and realized volatility in cryptocurrency markets. The MF-ADCCA approach revealed that the price-volatility relationship exhibits power-law cross-correlations as well as multifractal properties. We discussed the multifractal characteristics of the asymmetric cross-correlations when the market is rising and falling, together with the overall market trend, and found that the bivariate series have different properties between positive and negative market trends. We investigated the multifractal features in detail by using the generalized Hurst exponents and the singular spectrum. The results pointed out that generally for the investigated cryptocurrencies, cross-correlations of price and volatility in the uptrend markets have slightly higher persistency compared to those in the downtrend markets, irrespective of small and large fluctuations. Distinctive features of how small and large fluctuations operate on multifractality were also discovered and reported by investigating the spectrum distortion for each cryptocurrency and its market trends.

More importantly, the level of the asymmetric cross-correlations for each cryptocurrency was quantitatively evaluated by employing the asymmetric DCCA coefficient. Our empirical findings showed that depending on market directions or trends, the level of cross-correlations differs. We found the presence of stronger cross-correlations in bear markets than in bull markets for the maturing major coins (BTC and ETH), whereas the opposite results were observed for the still-developing minor coins (XRP and LTC). As long as price-volatility is our subject, we provided evidence that such an approach enables us to discuss whether asymmetric/inverse-asymmetric volatility dynamics are present with various time scales, which is an intriguing financial phenomenon for investors and financial regulators. The detection of asymmetric volatility works well since the results of our fractal analysis were in line with those of the conventional asymmetric GARCH-class models. Taking the advantage of the multifractal features, power-law cross-correlations, and scaling behaviors of price-volatility within various time scales, our approach can be an alternative approach for discussing dynamical volatility behaviors in cryptocurrency markets.

\begin{acknowledgments}
This research did not receive any specific grant from funding agencies in the public, commercial, or not-for-profit sectors.
\end{acknowledgments}

\appendix*
\section{stationarity test for volatility series}
\begin{table}[htbp]
\caption{\label{tab:stationarity_test}
Augmented Dickey-Fuller (ADF) test and Kwaitkowski-Phillips-Schmidt-Shin (KPSS) test for the logs of realized vipower volatility series ($\hat{\sigma}_t$). Notes: (1) The null is the existence of a unit root for the ADF test. (2) The null is stationarity for the KPSS test. (3) *, **, and *** represent the 10\%, 5\%, and 1\% significance levels, respectively.
}
\begin{ruledtabular}
\begin{tabular}{lllll}
 &\multicolumn{2}{l}{ADF test statistic}&\multicolumn{2}{l}{KPSS test statistic} \\
 \cline{2-5}
 $\hat{\sigma}_t$ & Level & Difference & Level & Difference \\
 \hline
 $\mathrm{BTC}$ & $-0.863$ & $-26.636^{***}$ & $1.728^{***}$ & $0.027$ \\
 $\mathrm{ETH}$ & $-0.306$ & $-11.455^{***}$ & $2.442^{***}$ & $0.024$ \\
 $\mathrm{XRP}$ & $-1.519$ & $-26.997^{***}$ & $0.752^{***}$ & $0.050$ \\
 $\mathrm{LTC}$ & $-1.260$ & $-24.639^{***}$ & $1.049^{***}$ & $0.045$ \\
\end{tabular}
\end{ruledtabular}
\end{table}

\nocite{*}

\bibliography{apssamp}

\end{document}